\documentclass[onecolumn,floatfix,aps]{revtex4}
\usepackage{amsfonts,amsmath,amssymb,amsthm}
\usepackage{graphicx,epsfig,hyperref} 
 
\begin{document}
\title{Fibonacci schemes of fault-tolerant quantum computation}

\author{Panos Aliferis}
\affiliation{IBM Watson Research Center, P.~O. Box 218, Yorktown Heights, NY 10598
            }
\begin{abstract}
The threshold estimate derived in previous versions of this paper was incorrect; this note explains the flaw. A new proof is discussed in arXiv:0809.5063.
\end{abstract}

\maketitle

As in earlier proofs of the quantum threshold theorem, previous versions of this paper \cite{fibonacci-old} analyzed a scheme for fault-tolerant quantum computation which had a self-similar structure: logical qubits were encoded using a concatenated code, and encoded gates were realized by gadgets which were constructed {\em recursively}. The code block at level $j{+}1$ of the concatenation hierarchy was constructed recursively using logical qubits encoded at level $j$, where the 4-qubit code $C_4$ was used at the first level. Similarly, the level-$(j{+}1)$ gadgets were built by replacing each fundamental gate in the level-$j$ gadget by the corresponding level-1 gadget, where the level-1 gadgets were constructed following Knill \cite{Knill05}. Because these level-1 gadgets were especially compact and simple, the intuition was that this scheme would yield a significantly higher threshold estimate than earlier proofs. 

The analysis of this scheme required different methods that previous analyses of recursive schemes because $C_4$ is a distance-2 code which cannot correct errors in unknown locations. As suggested by Knill \cite{Knill05}, there are two possibilities for how the scheme can operate: We can either use postselection accepting only those cases when the syndrome is trivial at every concatenation level, or we can use a message-passing decoding algorithm such that detected errors at level $j$ of the concatenated code become {\em located} errors that can be corrected at level $j{+}1$. The scheme based on postselection was analyzed numerically by Knill \cite{Knill05}, and later analyzed rigorously in \cite{Reichardt06,Aliferis07b}. The goal of this paper was to rigorously analyze the scheme based on message-passing decoding, and to support Knill's numerical findings \cite{Knill05} that it has an accuracy threshold comparable to the postselection scheme (whose disadvantage is, of course, the extreme overhead cost incurred by the extreme use of postselection).

To analyze the probability of failure of the scheme which uses message-passing decoding, we can consider separately each {\em flag history} --- that is, each possible set of subblocks which have been flagged as possibly having located errors in each particular run of the entire noisy quantum computation. Previous versions of this paper \cite{fibonacci-old} show is that if the physical noise strength of local stochastic noise is below $1.0 \times 10^{-3}$ then, for every flag history $h$, the {\em joint} probability of failure at level $j$ and realizing the history $h$ becomes arbitrarily small with increasing $j$.

It may appear that having such a result separately for every flag history, and in particular for the worst-case flag history, is sufficient. However, this is clearly not so as the following example illustrates: Suppose next to our quantum computer we have some unbiased coins, and we always flip them before each particular quantum computation (but otherwise nothing on the quantum computation is conditioned on the flip outcomes). We can modify our analysis so that except for all possible flag histories we also consider all possible {\em coin histories} --- that is, the outcomes of the coin flips. Then, for every flag history and for every coin history, we can surely make the joint probability of failure of the quantum computation and realizing these histories arbitrarily small by flipping a larger and larger number of coins. But this tells us nothing useful about the accuracy of the quantum computation, since we assumed that the coins are completely independent of everything else. The quantity one needs to consider is not the joint probability above but either $(a)$ the probability of failure {\em conditioned on} the worst-case  history, or $(b)$ the probability of failure obtained by summing over all possible histories. 

Calculating useful upper bounds on the conditional probability of failure is technically difficult because local stochastic noise allows faults to be adversarially correlated both temporally and spatially. Some progress could possibly be made by considering a less adversarial noise model and by using techniques similar to those in \cite{Aliferis07b}, but it seems that even then it would not be possible to prove a high threshold. On the other hand, calculating upper bounds on the probability of failure for all possible histories added together would be easier; it would require a new combinatorial analysis, but otherwise the recursion equations would be the same as in \cite{fibonacci-old}. But in this case also, it seems that the threshold estimate one would be able to prove would be significantly below ${\it O}(10^{-3})$. 

How can then one explain Knill's numerical findings? In fact, the scheme Knill analyzed is not only based on message-passing decoding, but it also uses clever gadget constructions which are {\em not} strictly recursive. Because gadgets at various coding levels are not self-similar, new methods of analysis are required and the proof of the threshold theorem for this scheme has very different structure. This non-recursive ``Fibonacci scheme''  is analyzed in \cite{fibonacci-new} which supersedes this paper.

I thank Barbara Terhal for the example with the coins discussed above.

\bibliographystyle{unsrt}

\end{document}